\documentclass[12pt,letterpaper]{article}

\usepackage[left = 1 in, right = 1 in, top = 1 in, bottom = 1 in]{geometry}

\usepackage{amsmath}
\usepackage{amssymb} 
\usepackage{graphicx}
\usepackage{setspace}
\usepackage{float}        
\usepackage{siunitx}
\usepackage{xspace}
\usepackage{textcomp}
\usepackage[dvipsnames]{xcolor}
\usepackage{listings}
\usepackage{url}
\usepackage[colorlinks,urlcolor=blue,linkcolor=blue,citecolor=blue]{hyperref}

\usepackage[superscript,biblabel]{cite}

\usepackage[symbol]{footmisc}

\newcommand{\reffig}[1]{Fig.~\ref{#1}}

\newcommand{\code}[1]{\mbox{\lstinline{#1}}}

\begin{document}
\sloppy 







\title{PHIDL: Python CAD layout and geometry creation for nanolithography}

\author{A. N. McCaughan$^1$, A. M. Tait$^1$, S. M. Buckley$^1$,\\ D. M. Oh$^2$, J. T. Chiles$^1$, J. M. Shainline$^1$ \& S. W. Nam$^1$}

\date{
    \small
    $^1$National Institute of Standards and Technology, Boulder, CO 80305\\%
    $^2$University of Colorado, Boulder, CO 80305\\%
}
\maketitle

\pagenumbering{arabic} 

\section*{Introduction}

Computer-aided design (CAD) has become a critical element in the creation of nanopatterned structures and devices. In particular, with the increased adoption of easy-to-learn programming languages like Python there has been a significant rise in the amount of lithographic geometries generated through scripting and programming. However, there are currently unaddressed gaps in usability for open-source CAD tools -- especially those in the GDSII design space -- that prevent wider adoption by scientists and students who might otherwise benefit from scripted design. For example, constructing relations between adjacent geometries is often much more difficult than necessary -- spacing a resonator structure a few micrometers from a readout structure often requires manually-coding the placement arithmetic. While inconveniences like this can be overcome by writing custom functions, they are often significant barriers to entry for new users or those less familiar with programming. To help streamline the design process and reduce barrier to entry for scripting designs, we have developed PHIDL\footnote[2]{\url{https://github.com/amccaugh/PHIDL}}, an open-source GDSII-based CAD tool for Python 2 and 3 based on gdspy\cite{Gabrielli} and numpy\cite{VanderWalt2011}.

In PHIDL, we have placed an high priority on usability, clarity, and consistency: the package is purpose-built so that a brand-new user can learn the conceptual underpinnings and begin designing useful geometries in just a few minutes. In its development, we sought to emulate the ease of vector-editing software like Inkscape and Adobe Illustrator as these programs and their interfaces have been developed for decades and are highly intuitive even to new users.

At present, several other geometry creation CAD tools do exist, with varying levels of scripting interfaces \cite{Gehring2019,Balram2017,Klofferlein,Bogaerts2012a}. Generally, the software built for scripting GDS geometries follow the GDS specification closely, including polygons, cells, references, and arrays. However, PHIDL also provides functionality for the user well beyond just the GDS specification.

\begin{figure}[h] 
\centering
    \includegraphics[width=3in]{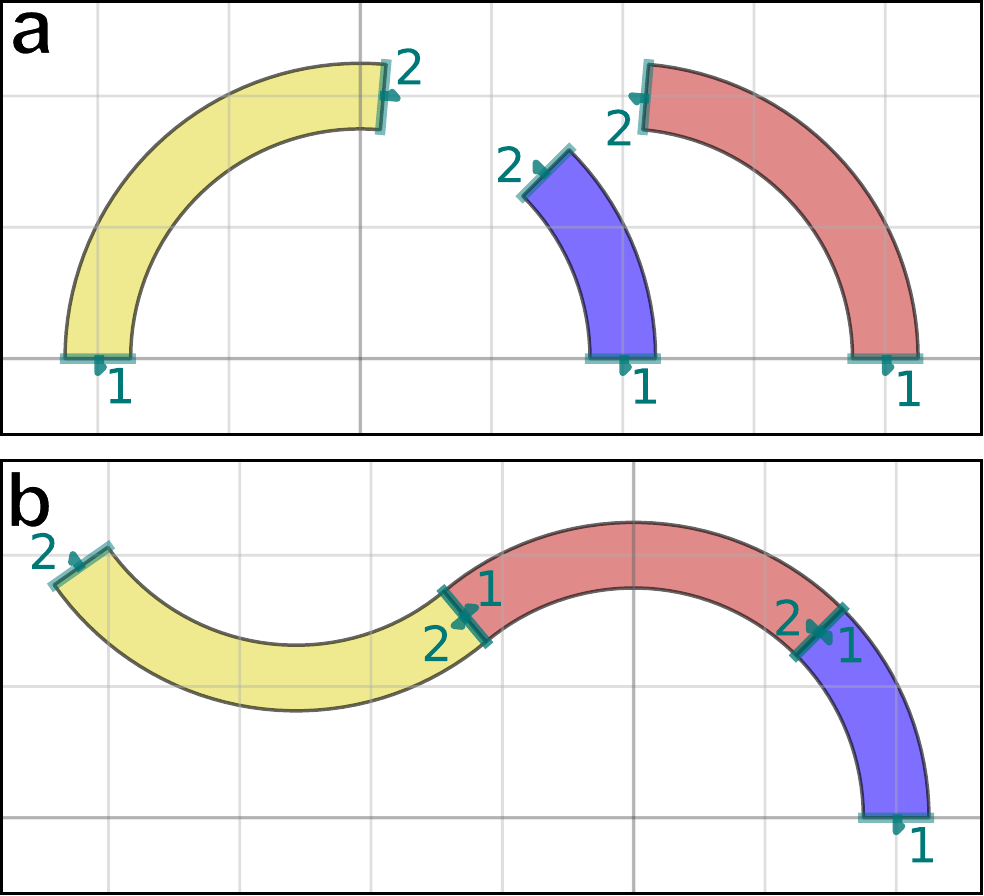}
    \caption{Creating complex geometries by connecting shapes together. (a) Three arcs are generated, and each arc has a ``port'' on either end. (b) The shapes are snapped together like building blocks using the \code{connect()} function, which automatically performs the calculations needed to mate the ports together.}
    \label{ports}
\end{figure}

The basic premise of PHIDL is that polygons are created by the user and grouped into one or more ``Device'' objects. (For those familiar with GDS design, a Device is just a ``cell'' with a few special features). Within the Devices, a user can also define ``ports'' which offer a convenient way of connecting one Device to another--ports are typically placed at the input and outputs of a Device.  For example, when building a smooth path with contacts at either end, a user would likely put ports on the path outputs and, separately, also put ports on the contacts. After a few Devices are constructed by the user, they can be trivially snapped together like building blocks using the \code{connect()} function.

\begin{figure}[h] 
\centering
    \includegraphics[width=3in]{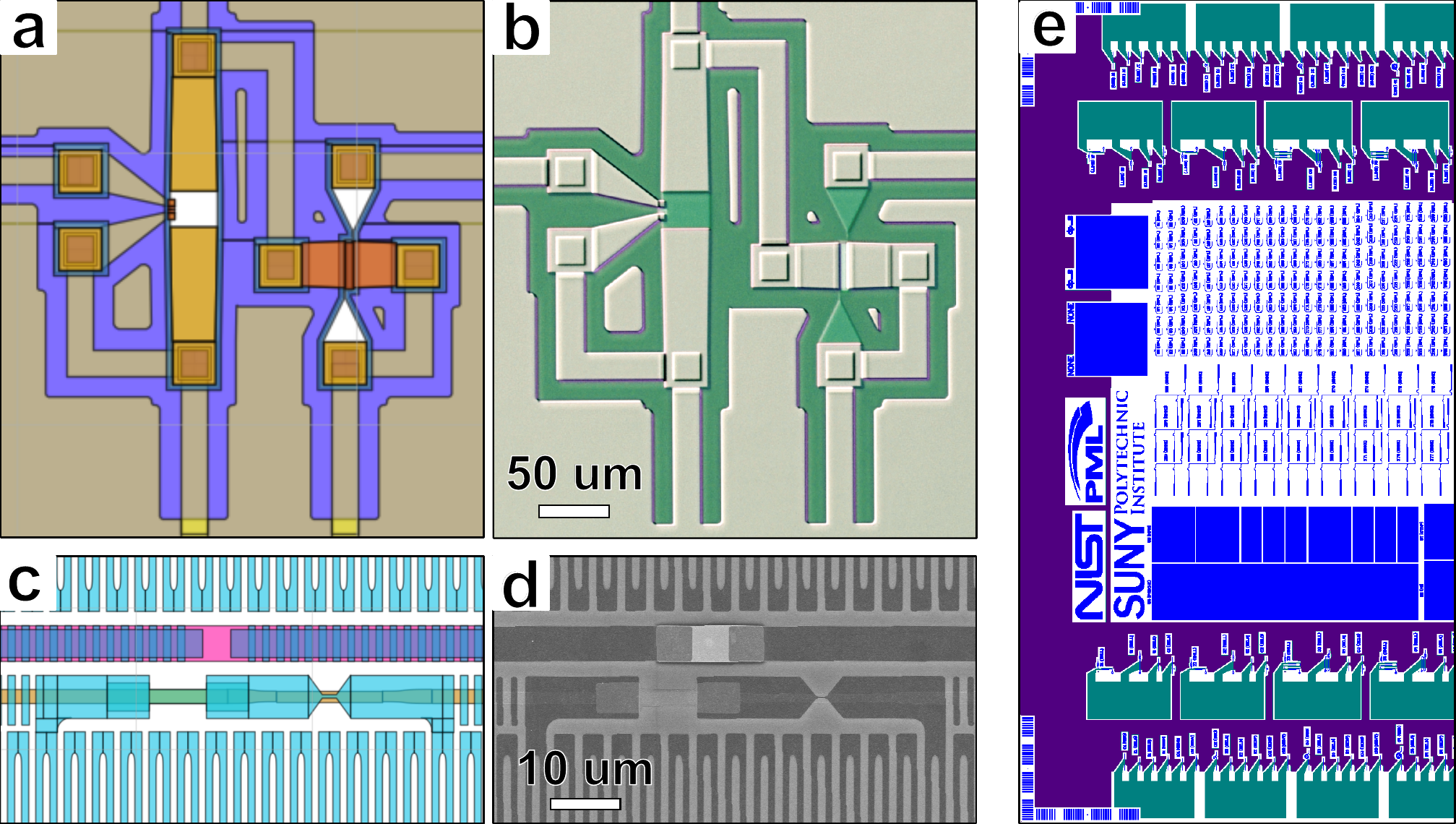}
    \caption{Example layouts and fabricated die made with PHIDL (a) Layout of a superconducting-nanowire electronic circuit based on thermal switches\cite{McCaughan2019b} with 7 independent layers, and (b) resulting microscope image of fabricated device. (c) Layout of a multilayer superconducting single photon detector device and (d) scanning electron micrograph of the fabricated device. (e) Photonic integrated circuit layout made in PHIDL with over 95,000 polygons and 6.8 million points.}
    \label{fabricated_designs}
\end{figure}

In this way, very complex geometries can be assembled one piece at a time without requiring the user to manually compute placement locations. We note that this kind of shape-to-shape snapping is ubiquitous in vector-editing and other design software (e.g. Adobe Illustrator\texttrademark, Inkscape, and even Microsoft PowerPoint\texttrademark) due to its convenience and utility. The Device also makes a convenient abstraction for working with complex polygons, as the user does not need typically to concern themselves with the details of the polygons inside the Device--only use the ports to quickly connect it to other Devices. Shown in \reffig{fabricated_designs} are photonic and superconducting layouts made in PHIDL which were successfully fabricated in a cleanroom.

\begin{figure}[h] 
\centering
    \includegraphics[width=3in]{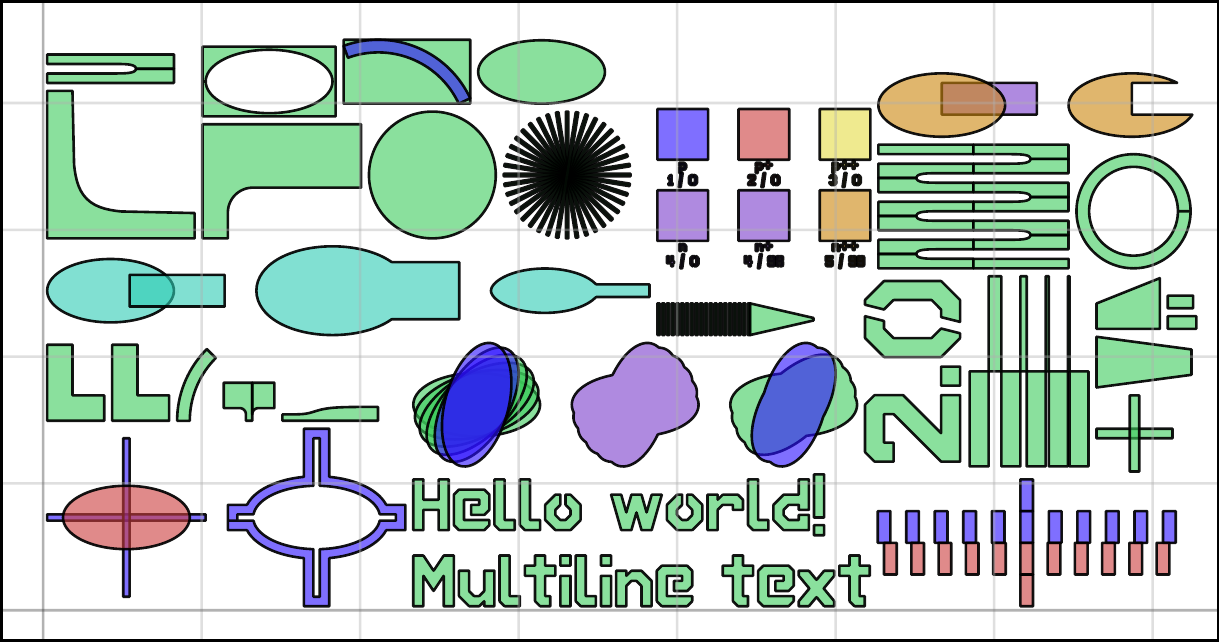}
    \caption{Examples of the built-in geometry library functions.These include basic shapes, text, layer-alignment calipers, resolution tests, boolean operations, and grow/shrink operations. These shapes were automatically placed within the rectangular area using PHIDL's built-in packing algorithm \code{packer()}.}
    \label{library}
\end{figure}

To make PHIDL as convenient as possible, there is an included geometry library \code{phidl.geometry} (known as \code{pg} hereafter) which contains a large number of easy-to-use functions which can produce basic shapes (ellipses, rectangles, arcs, etc), advanced shapes (text, contact pads, etc), boolean functions (boolean, union, offset, etc), lithographic test structures (resolution tests, calipers, etc) and application-specific shapes such as photonic waveguides structures and superconducting nanowire single photon detectors. One particularly useful feature is the \code{packer()} function, which takes a list of shapes or Device objects and packs them together into as small an area as possible.  Shown in \reffig{library} is the usage of this function to automatically pack all of the built-in geometry library shapes into a single area.

This package also includes a variety of other useful features, such as (1) the \code{quickplot()} function which can quickly generate a plot of any shape or device using matplotlib\cite{Hunter2007}, (2) the ability to easily to record custom metadata for every Device and export it for later reference, and (3) the ability to export designs directly to the SVG format for use in figures and scientific posters.

\section*{Design concept overview}

This software package is aimed at scientists and students who need to manipulate geometry and produce GDS files and want to do that a minimal amount of time spent learning programmatic structure.  Below, we enumerate the principles according that this package strives towards, listed in order of importance.

\subsubsection*{Usability}

The goal in designing PHIDL was to ensure that anyone who has used common drawing software can start designing useful structures with as little start-up time as possible. We found that many of the current GDS scripting methods encourage users to position geometry according a ``rubber stamp'' approach that consists of:  (1) selecting a geometry function (e.g. a ring or a transistor), (2) choosing the exact coordinates for its location, then (3) creating an instance of the geometry at that position. Much like a mark made by a rubber stamp, if the position of the geometry is found to be incorrect later it is not easy to move it after the fact; the user must trace back through the code, modify the position of the geometry at its instantiation, and update any follow-on calculations which depend on that positioning. For PHIDL, we opted to take the ``building block'' approach used by many types of graphical-editing software, where the user is encouraged to gather structures together first and then assemble them afterwards.  In PHIDL, this translates to (1) creating instances needed geometry immediately, not worrying about positioning, then (2) manipulating the geometries together with  \code{move()}, \code{rotate()}, \code{reflect()}, and \code{connect()} until everything is positioned as as desired. We note that at very large scales there are useful optimizations possible with the ``rubber stamp'' approach--however the optimizations are not large compared against the overhead of Python itself, and we believe the usability and intuitiveness of our approach make for much larger time savings overall.

An important aspect to making the ``building block'' approach usable was to guarantee that any PHIDL object can be manipulated/transformed using the same key words (e.g. \code{move()} or \code{rotate()})--whether it be the geometry-containing Device itself, a reference to that Device, a polygon, or a port.  This consistency reduces the amount of memorization required by the user and follows the principle of least astonishment, which states that ``...a component of a system should behave in a way that most users will expect it to behave; the behavior should not astonish or surprise users.'' Aiming for a high level of usability also meant we put a priority on having many examples. Documentation always useful, but our experience has been that examples are the quickest way to start using scripted software, with documentation acting as a canonical resource for deeper understanding of the functions. To this end, we have created a wealth of examples available as tutorials and in the online documentation.

\subsubsection*{Flexibility}

Similar to usability, PHIDL is also designed to be flexible in the ways that the user can interact with and manipulate its objects. This means that PHIDL was designed with the software principle of ``do what I mean'' which states that the user ``...should not be stopped and forced to correct themselves or give additional information in situations where the correction or information is obvious.'' An example of following this principle is that polygons can be created by entering the data either as a list of x/y pairs [(x1,y1),(x2,y2),(x3,y3),...] or as a pair of ordered lists [(x1,x2,x3,...),(y1,y2,y3,...)]. PHIDL will deduce which format the user's data is being entered in and make the necessary conversions without user intervention. (Since any non-trivial 2D polygon must have at least 3 points there can be no ambiguous cases.) Another example of flexibility is how PHIDL allows the user to position geometry in multiple ways.  As a first option, the user can manually move an object such as a Device or a polygon with its \code{move()} command and the related fixed-axis commands \code{movex()} and \code{movey()}.  As a second option, the user can use relational properties of the geometry to move objects around.  For instance, to separate two circles in the x-direction by exactly 5 units one can use the command \code{circle1.xmin = circle2.xmax + 5}. This will position \code{circle1} such that its minimum x value is 5 units to the right of the maximum x value of \code{circle2}. As a third option, the \code{connect()} command can be used to snap together Devices like building blocks using their ports. The \code{connect()} command even includes an optional \code{overlap} argument which can be used force overlap between geometries.  This overlap is often useful in multi-layer cleanroom fabrication when perfect alignment between layers cannot be guaranteed.

Also useful is the flexibility in layer specification.  To comply with GDS standards every polygon must have a \code{layer} and \code{datatype} -- PHIDL follows this convention, but allows more flexible entry in specifying the layer.  For example, making cross shape on layer 7 with datatype 0 can be accomplished either by explicity writing \code{pg.cross(layer = (7,0))} or using the shorthand \code{pg.cross(layer = 7)}. When creating geometry, multiple layers can also be specified easily using the built-in Python \code{set} object.  For example using the argument \code{layer = \{7, 8, (9,25)\}} will put copies of the geometry on layers \code{(7,0)}, \code{(8,0)} and \code{(9,25)}. Alternatively, there is a more advanced PHIDL \code{LayerSet} object which can be used to group several layers together. There is also an 'alias' functionality within phidl.  To work with an object, it can either be stored into its own variable or it can be assigned to the Device which owns it with an 'alias'.  For instance, if we want to add a reference of a circle to a Device \code{D} using the \code{add\_ref()} function, we can either hang on to the reference by (1) assigning it to a temporary variable like \code{myrect = D.add\_ref(E)} or by (2) assigning it an alias within \code{D} like \code{D[`myrect'] = D.add\_ref(E)}.  Storing objects as aliases can also help reduce ambiguity when sharing code because it becomes obvious to which Device the object belongs.

\subsubsection*{Minimal structure}

When designing PHIDL, we tried to create as simple a software structure as was feasible while still prioritizing usability. We have observed that scientists and graduate students are often part-time designers, using design tools heavily for a few days or weeks then leaving the tools for a period of time to implement the designs. As a result, we tried to keep the structure of PHIDL minimal so that both a first-time user or a returning user has a minimal amount of architectural information they have to keep memorized. One example of keeping structure to a minimum is that all functions in the \code{phidl.geometry} library operate the same way: they create new geometry and return a single Device.  Similarly, all functions within a Device object (for instance, \code{align()} or \code{flatten()}) only modify the existing geometry within the Device.

We also attempted to provide sane defaults for any function that has potentially ambiguous arguments. These defaults give the user a starting point, and by viewing the resulting geometry with the \code{quickplot()} function the user can learn-by-inspection. This process can occur without ever leaving the Python terminal, needing to read the function code, or looking up the the online documentation. PHIDL also offers convenience functions wherever possible.  For example, while it is possible to align several polygons horizontally along the y-axis using the \code{move()} command and a \code{for} loop, we have also included a convenience function in the Device class called \code{align()} because it is such a common action in geometry design.

Without enumerating them all, other convenience functions include the \code{distribute()} function (distributes a list of geometries so they have fixed spacing between them), movement-related command related to the bounding box mentioned earlier ({xmin}/{ymax}/etc), and the \code{packer()} function shown in \reffig{library} (packs geometries into the smallest area possible). These single-line convenience functions increase the readability of the code and are so useful they are implemented in virtually all vector-editing software. Lastly, PHIDL attempts to use pre-existing conventions from related software where possible. Table~1 shows the variations of nomenclature used in related geometry-editing software (including GDS editing software and vector-editing software).  For the sake of user ease, PHIDL attempts to follow the most widely-used conventions.

\begin{table}[]
\centering
\footnotesize
\caption {Naming conventions of geometry-editing software}
\label{tab:title} 
\begin{tabular}{l|cccc}
               & Reflection      & Translation   & Rotation        & Offset          \\ \hline
\textbf{PHIDL} & \textbf{mirror} & \textbf{move} & \textbf{rotate} & \textbf{offset} \\
gdspy\cite{Gabrielli}          & mirror          & translate     & rotate          & offset          \\
IPKISS\cite{Bogaerts2012a}         & mirror          & move          & rotate          & offset/grow     \\
nazca\cite{Broeke}          & flip            & move          & rotate          & buffer          \\
gdshelpers\cite{Gehring2019}     & transform       & translate     & rotate          & buffer          \\
KLayout\cite{Klofferlein}        & mirror          & move          & rotate          & size            \\
LayoutEditor\cite{Thies}   & mirror          & move          & rotate          & sizeadjust      \\
L-edit         & mirror          & move          & rotate          & grow            \\
Illustrator    & reflect         & move          & rotate          & offset          \\
Inkscape       & flip            & move          & rotate          & offset          \\
Powerpoint     & flip            & move          & rotate          & N/A            
\end{tabular}
\end{table}

\section*{Key elements of geometry library}

Here we list a few of the pre-made geometries available in the built-in \code{phidl.geometry}.  A more complete list can be seen in the online documentation\footnote[2]{\url{https://phidl.readthedocs.io/}}.

\subsubsection*{Basic shapes}

These are shapes which were included due to their widespread utility.  They include functions such as \code{rectangle()}, \code{arc()}, \code{circle()}, \code{ellipse()}, \code{ring()}, and more. We note that although a circle is a subset of an ellipse, it is also one of the most commonly used shapes, so to match user expectations \code{circle()} is provided as shortcut to the \code{ellipse()} function. Similarly there are rectangle variants such as \code{bbox()} that can be used to easily draw a bounding box around an object, and \code{compass()} which is a rectangle with ports placed on each edge. We note that geometry-creating functions in PHIDL don't have arguments to specify the position--this is done to avoid ``rubber-stamp'' style design.  Additionally, these arguments are often ambiguous -- for example, if a rectangle-generation function has a \code{position} argument, it will be unclear whether that position refers to the rectangle center or one of its corners. To facilitate code clarity, PHIDL encourages users to perform operations like centering after creation of the geometry (e.g. \code{my\_rectangle.center = (10,5)}) The \code{text()} function is another basic one which can print multiline text with left, right, or center justification.  In addition to having full unicode font support, it also includes the DEPLOF font, which was specially designed for photolithography and electron-beam lithography to avoid islands/delamination of resist.

\subsubsection*{Paths / waveguides}

\begin{figure}[h] 
    \centering
        \includegraphics[width=3in]{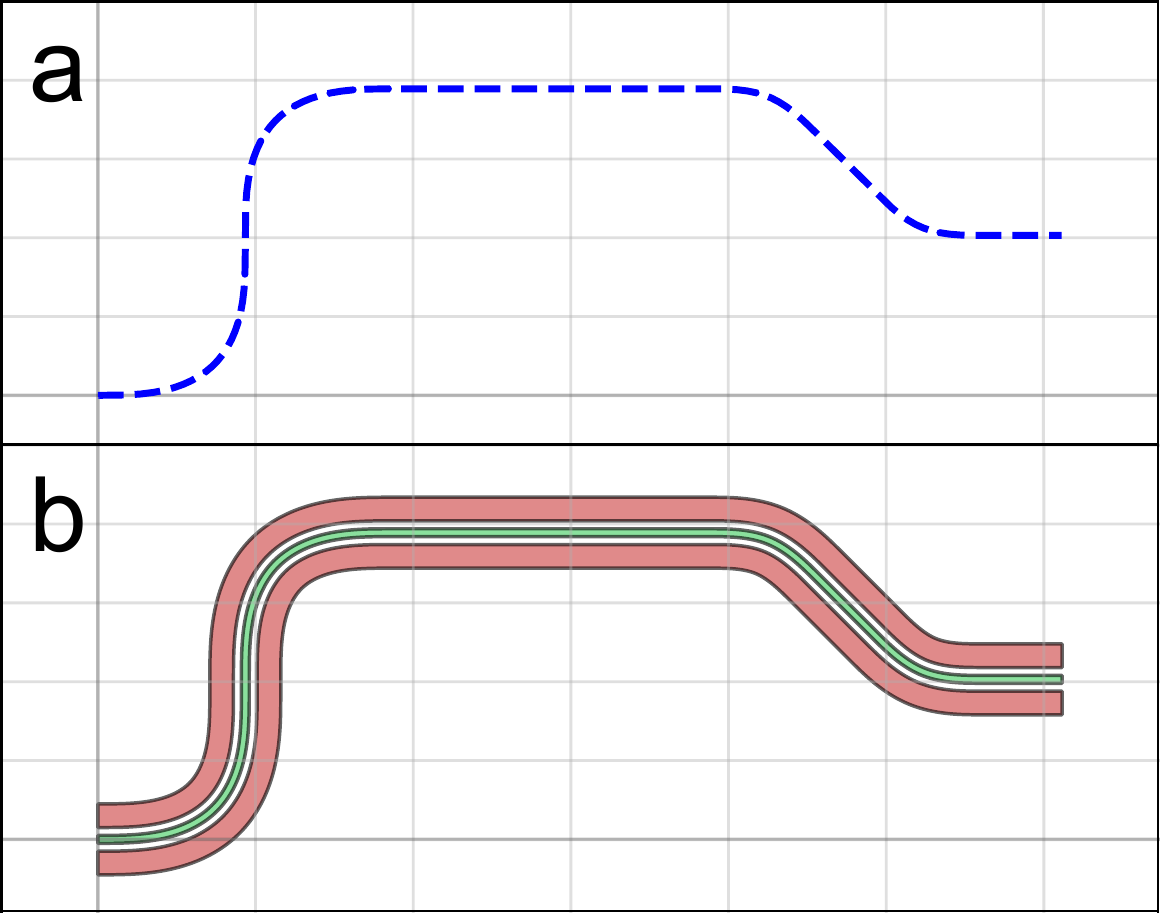}
        \caption{Path / waveguide module in phidl. (a) Construction of a path from circular \code{arc()} sections, \code{straight()} sections, and straight-to-bend \code{euler()} sections.  (b) Combining the 1D path with a 1D cross-section to create a set of 2D polygons.}
        \label{waveguide}
    \end{figure}

The package also includes a highly efficient module for creating smooth curves, particularly useful for creating waveguide structures such as those used in photonics.  The process is designed to be intuitive and powerful, by conceptually separating the specification of the path from the specification of the cross-section.  The path can be constructed piece by piece using the \code{append()} functionality and several convenient built-in component functions (such as the \code{arc()} section, the \code{straight()} section, or the straight-to-bend \code{euler()} curve section, also known as a track transition or clothoid).  Separately, the cross-section can be defined in a similar manner. By combining the 1D path and the 1D cross-section, the final 2D polygons can be easily output as shown in \reffig{waveguide}.  PHIDL also includes a fast implementation of the Ramer-Douglas–Peucker algorithm\cite{Douglas1973} for polygon simplification, as one of the chief concerns of generating smooth curves is that too many points are generated, inflating file sizes and making boolean operations computationally expensive.  The path module also comes with a built-in \code{transition()} function that allows simple transitioning between two cross-sections along an arbitrary Path.

\subsubsection*{Boolean / offset}

\begin{figure}[h] 
\centering
    \includegraphics[width=3in]{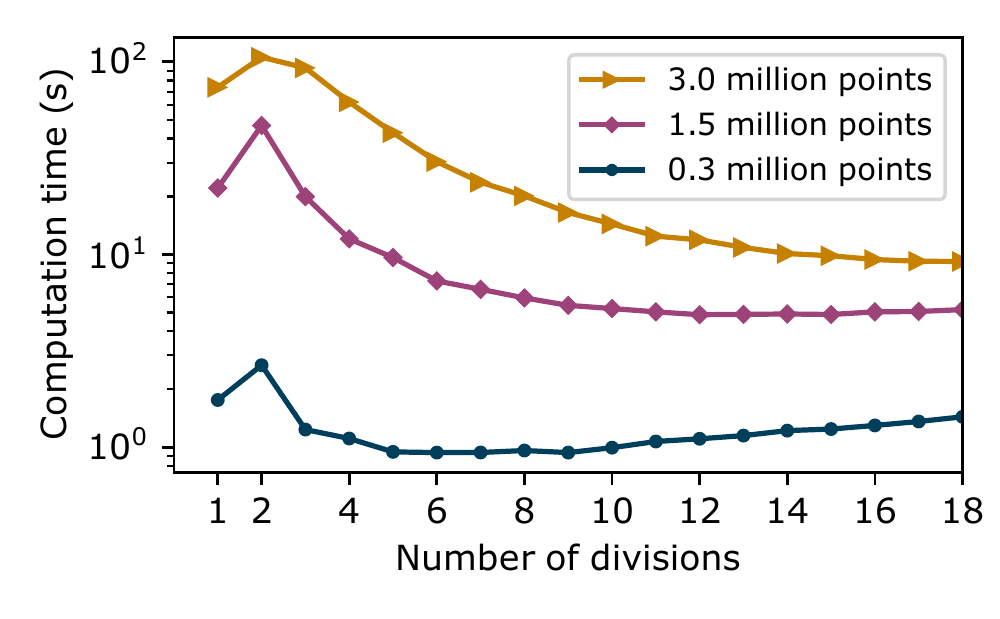}
    \caption{Speedup of boolean operations by subdivision. Shown is the effect of varying the \code{num\_divisions} parameter when performing the \code{boolean()} on a large number of random shapes.  When $n$ is greater than 1, the geometry is partitioned into $n \times n$ equal-sized rectangles and the boolean operation is applied to each subdivision sequentially, enabling significant speedup of the operation. }
    \label{offset_subdivision}
\end{figure}

The \code{boolean()} function can perform the standard suite of AND (intersection), OR (union), NOT (subtraction), and XOR (exclusive disjunction) operations, and the \code{offset()} function provides grow/shrink operations for polygons. The codebase for this is provided by the Clipper library\cite{Johnson} by way of \code{gdspy}. Performing boolean and offset operations on 2D geometry can be computationally intensive, so we have added functionality to optimize those operations.  By setting the  \code{num\_divisions} argument in \code{boolean()} and \code{offset()}, the user can choose to slice the geometry into multiple subsections before performing the operation.  Since the boolean and offsetting operations are generally more complex than $O(N)$ for $N$ points, dividing the geometry into multiple sections can speed up the operation significantly, as shown in \reffig{offset_subdivision}. This process enables significant speedup, but since the the subdivision process adds computational overhead, too-small or too-large $n$ can result in increased computation time.

\subsubsection*{Lithographic test structures}

Although lithographic tests used in cleanroom fabrication are often application-specific, we found that easy access to a few of the most common structures has been critical to encourage users to include lithographic tests in their designs. Among these test structures include \code{litho\_steps()} (tests linewidth and resolution), \code{litho\_star()} (tests linewidth and aliasing), and \code{litho\_calipers()} (checks alignment accuracy between two layers). We have also implemented a filling function \code{fill\_rectangle()}, which can be used to populate empty areas of a wafer with configurable-density rectangles for more uniform photoresist development.

\subsubsection*{Application-specific geometries}

There are several geometries in PHIDL which are meant for superconducting and photonic geometry creation.  Included are functions like \code{snspd()}, for creating superconducting nanowire single-photon detectors using optimized curves which reduce the problem of superconducting ``current crowding''. There are also test structures such as \code{test\_ic()} For photonic creation, the \code{phidl.path} functions allows users to create complex paths and waveguides by specifying (1) a path of points that the path should follow and (2) the 1D cross-section of the path. There are also functions meant for automatic routing between paths.  For example, \code{route\_basic()} can connect the ports of two paths together using a smooth sine curve. For more advanced routing, \code{route\_manhattan()} will smoothly connect paths together along a manhattan grid.

\subsubsection*{User-defined geometries}

Of course, the user can always create their own geometry functions within phidl. In this process, users are encouraged to make use of the same style as the phidl.geometry library: (1) the function should return only a single Device (2) metadata about the geometry should be saved in the Device's \code{.info} variable, which is a Python dictionary meant for saving (and later retrieving) information about the geometrical object. If the user finds themselves designing geometries which take a large amount of time to compute, they can use the \code{@device\_lru\_cache} decorator, which allows caching of geometries so they only have to be calculated once per set of arguments.

\section*{Conclusion}

In summary, PHIDL is geometry manipulation tool aimed at scientists, graduate students, and anyone trying to script the creation of 2D geometries. Like Python itself, it aims to be readable, and intuitive. To this end, the software design focuses on usability, flexibility, and simplicity. The goal has been to develop a GDSII-creation tool which can be picked up by new or returning users in a few minutes, allowing users to get to desiging as quickly as possible while maintaining an architecture robust enough to build extremely complex geometries. It comes with a large library of premade geometry functions and many convenience functions for the creation, manipulation, and debugging of large scripted geometries. There are also opportunities to further optimize PHIDL, for instance using a compiled backend (such as KLayout\cite{Klofferlein}) as a faster geometry database, or using just-in-time compilation packages such as \code{numba}\cite{Lam2015} for computationally-expensive operations.

The U.S. Government is authorized to reproduce and distribute reprints for governmental purposes notwithstanding any copyright annotation thereon. Certain software and commercial materials are identified in this paper to foster understanding. Such identification does not imply recommendation or endorsement by the National Institute of Standards and Technology, nor does it imply that the materials or equipment identified are necessarily the best available for the purpose.





\newpage
\bibliographystyle{unsrt}

\end{document}